%% file: main.tex
\documentclass[conference]{IEEEtran} %
\usepackage{graphicx} % Required for inserting images
\usepackage{xcolor}

\usepackage{subcaption} %
\usepackage{adjustbox} %
\usepackage{listings}
\usepackage{hyperref }
\input{lsts/config-default} %
\input{lsts/config-mlir} %

\date{January 2024}
\begin{document}

\title{Data Transfer Optimizations for Host-CPU and Accelerators in AXI4MLIR}

\input{IEEE/authors}

\maketitle 

\begin{abstract}
As custom hardware accelerators become more prevalent, it becomes increasingly important to automatically generate efficient host-driver code that can fully leverage the capabilities of these accelerators. This approach saves time and reduces the likelihood of errors that can occur during manual implementation. \textbf{AXI4MLIR}~\cite{agostini2023axi4mlir} extends the MLIR compiler framework~\cite{Lattner2021mlir} to generate host-driver code for custom accelerators for linear algebra problems. By leveraging specific compiler optimizations, we can further increase accelerator utilization. 

In this work we offer two key observations through a MatMul accelerator case study. First, the accelerator's compute core utilization is less than 10\%, and second, the critical latency bottleneck is caused by copying data between the heap and memory-mapped DMA buffers. We identify a set of missing host code optimizations to improve the under-utilization and the latency bottleneck. Therefore, we propose three key host-code data-movement-related optimizations, extending AXI4MLIR. The optimizations provide DMA-based data allocation, coalescing of DMA transfers, and pipelining of the accelerator's load, compute, and store stages.

\end{abstract}

% ***************************************************************

\section{Introduction}

Renewed interest in heterogeneous computing with custom hardware accelerators can be attributed to the diminishing performance improvements of traditional architectures, especially as the core count grows. Custom hardware accelerators provide increased performance through architectural-level support that is tailored to the specific requirements of the given problem or application ~\cite{shabani2023hpca,kim2023hpca,hsia2023asplos,munoz2023asplos}. 
Tensor algebra processing, pervasive in machine learning ~\cite{Rao2018sdc} applications, is computationally demanding and requires high memory bandwidth. Specialized accelerators are ideal for providing the computational power necessary to meet the needs of these applications. Leveraging
custom accelerators poses a new challenge - how do we realize their full potential, since it requires efficient communication between the accelerator, off-chip memory, and host CPU, demanding effective hardware-software co-design of the accelerator and host driver code. An efficient accelerator driver is necessary to handle data transfers, synchronize operations, and control the accelerator with instructions or control signals. However, manually developing host driver code for a particular accelerator and different applications is a very time-consuming and error-prone process.

% Description and figure of AXI4MLIR
To solve this problem, we can utilize automated CPU-accelerator driver code generation, which considers the host-CPU, the accelerator, the associated communication protocol, and the available accelerator instructions. \textbf{AXI4MLIR}~\cite{agostini2023axi4mlir} is an extension of the MLIR compiler framework~\cite{Lattner2021mlir} that generates host-driver code which offloads linear algebra operations to custom AXI-Stream based accelerators. AXI4MLIR takes a high-level application description in the MLIR’s linear algebra (\lstinline{linalg}) abstraction~\cite{mlir2020linalg} as input and introduces custom MLIR attributes to describe the target accelerator capabilities, which eventually gets transformed into low-level DMA library calls to send or receive data and instructions to the accelerator.

\begin{figure}[t]
    \centering
    \includegraphics[width=1\linewidth]{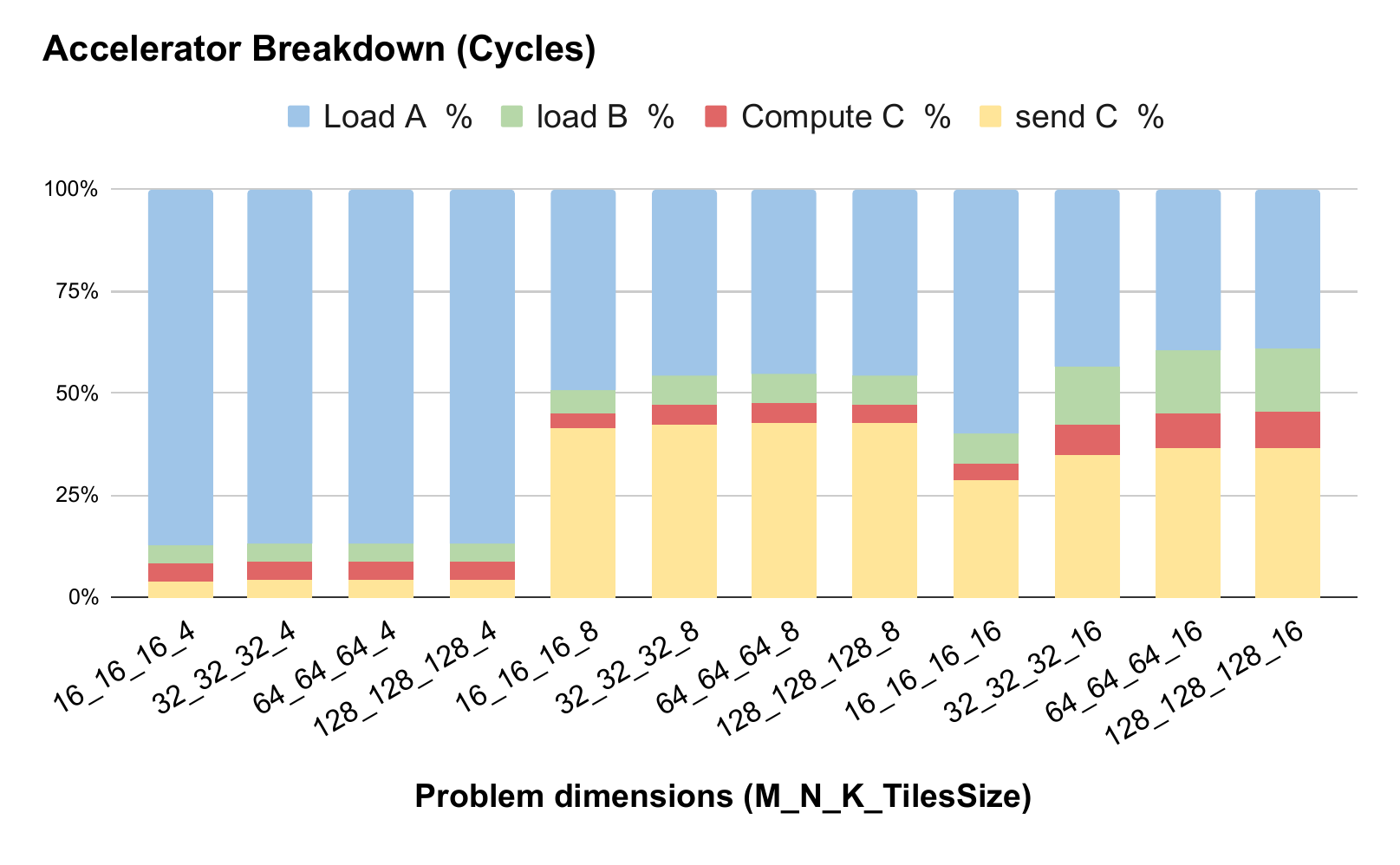}
    \caption{Breakdown of clock cycles spent inside a simple MatMul accelerator. Red segments (Compute C \%) represents the time where the accelerator's processing elements are active.}
    \label{fig:acc_breakdown}
\end{figure}

\begin{figure}
\centering
\begin{adjustbox}{minipage=\columnwidth,scale=0.9}\begin{subfigure}[t]{1\textwidth}
\centering
\input{lsts/baseline.tex}
\caption{Baseline code generation}\label{fig:baseline}
\end{subfigure}

\begin{subfigure}[t]{1\textwidth}
\centering
\input{lsts/dma-allocation.tex}
\caption{With DMA-based data-allocation.}\label{fig:dma-allocation}
\end{subfigure}

\begin{subfigure}[t]{1\textwidth}
\centering
\input{lsts/data-coalescing.tex}
\caption{With data-coalescing.}\label{fig:data-coalescing}
\end{subfigure}

\begin{subfigure}[t]{1\textwidth}
\centering
\input{lsts/pipeline}
\caption{With software pipelining of transfers.}\label{fig:pipeline}
\end{subfigure}

\end{adjustbox}
\caption{Psedo-MLIR code of a tiled MatMul problem showcasing baseline and proposed extensions.}
\label{fig:host-driver}
\end{figure}

% Problems and motivation for data optimizations
While AXI4MLIR can provide considerable speedup over a manual implementation of the host-driver code, we observe potential areas for further optimizations, which can improve the utilization of the accelerator and reduce overall latency. Figure~\ref{fig:acc_breakdown} shows a breakdown of the cycles spent inside the accelerator while executing MatMul problems for a range of dimensions and tile sizes. Ideally, the accelerator's compute cores should be fully utilized, but our experiments show that we are achieving on average less than 10\% utilization. Additionally, profiling the execution of this MatMul problem from the CPU's perspective, we observed that the CPU-side bottleneck is caused by copying data allocated within the memory heap to the DMA buffers. To tackle this accelerator under-utilization and host-side bottlenecks, we propose extending AXI4MLIR with three key optimizations, discussed in detail in the following section.

% ***************************************************************

\section{Proposed Data Transfer Optimizations}

The following data movement optimizations extend the AXI4MLIR transformation and lowering pipeline to mitigate and hide the time spent on transferring data.
We demonstrate the apply the proposed optimizations to the tiled MatMul problem with a flexible stream-based accelerator that supports double buffering. Figure~\ref{fig:baseline} shows the baseline code generated by AXI4MLIR for this accelerator and algorithm.

\subsubsection{DMA-based data allocation}

Any data that needs to be transferred via AXI DMA engines must first be placed in DMA-mmapped buffers.
Current AXI4MLIR implementations copy data between buffers (\lstinline{memrefs}) allocated in the heap to the mmapped region (dma buffers) while communicating with the accelerator. This can incur significant overhead. To mitigate the extra staging transfers, we propose a new attribute and set of transformations that trigger allocation of \lstinline{memrefs} needed by the accelerator directly in the DMA region. 

Figure~\ref{fig:dma-allocation} demonstrates how the optimization will tag the \lstinline{memref} with the \lstinline{#dma} tag, this ensures that the required data is allocated within the DMA buffers, hence avoiding additional copying. To support this feature, the lowering of \lstinline{accel.send} operations is simplified, skipping any staging copy and immediately initiating data transfers to the accelerator.

\subsubsection{Data coalescing}

A detailed analysis of the accelerator performance highlights that the initial data load latency is a bottleneck within the accelerator, which explains why \textit{load A} takes more cycles than \textit{load B} within Figure~\ref{fig:acc_breakdown}. The first data packet, loaded from memory to accelerator, incurs additional latency, whereas the following data can be loaded in a fifo-like manner with, in some cases, a 1-cycle delay. To reduce the initial transfer cost, we propose a data coalescing strategy. This involves combining multiple data sends within the same loop body into a single DMA transfer operation, replacing multiple synchronization operations with a single.

Figure~\ref{fig:data-coalescing} shows how we propose to represent the coalescing of multiple data transfers into one. We will transform \mbox{\lstinline{accel.send}} into a \textit{variadic} operation, which takes multiple \lstinline{memref} arguments, indicating that only one synchronization is necessary.

\subsubsection{Software pipelining \& double-buffering}

To further reduce the accelerator idle time and overall latency, we propose host-driver support to pipeline the accelerator load, compute, and store stages. Figure~\ref{fig:pipeline} demonstrates how the initial iterations of the inner loop within the \lstinline{matmul_call} are moved outside the loop to enable software-level pipelining. Note that this driver code optimization can only be enabled when the accelerator provides double buffering support to overlap loading data with computation.

% ***************************************************************

\section{Conclusion}

Manually writing host-driver code for custom accelerators is time-consuming and error-prone. AXI4MLIR provides a code-generation solution to generate efficient host-driver code. Here, we propose a set of data-related optimizations for  AXI4MLIR to improve accelerator utilization and overall latency.

% ***************************************************************

\bibliographystyle{IEEEtran} %
\bibliography{main}

\end{document}

%% file: lsts/config-default.tex
\definecolor{codegreen}{rgb}{0,0.6,0}
\definecolor{codegray}{rgb}{0.5,0.5,0.5}
\definecolor{codepurple}{rgb}{0.58,0,0.82}
\definecolor{backcolour}{rgb}{0.95,0.95,0.92}

\lstdefinestyle{mystyle}{
	backgroundcolor=\color{white},
	keywordstyle=\color{codegreen},
	numberstyle=\tiny\color{codegray},
	stringstyle=\color{codepurple},
	basicstyle=\ttfamily,
	breakatwhitespace=false,
	breaklines=true,
	captionpos=b,
	keepspaces=true,
	numbers=left,
	numbersep=5pt,
	showspaces=false,
	showstringspaces=false,
	showtabs=false,
	tabsize=2,
    morekeywords=[1]{
    },
}

%% file: lsts/config-mlir.tex
\lstdefinestyle{customMLIR}{
    inputencoding=utf8,
    tabsize=2,
    rulecolor=,
    upquote=true,
    columns=fixed,
    linewidth=\columnwidth,
    showstringspaces=false,
    extendedchars=true,
    breaklines=true,
    showtabs=false,
    showspaces=false,
    showstringspaces=false,
    basicstyle=\scriptsize\ttfamily,
    identifierstyle=\scriptsize\ttfamily,
    keywordstyle=\scriptsize\ttfamily\color[rgb]{0,0,1},
    commentstyle=\scriptsize\ttfamily\color[rgb]{0.133,0.545,0.133},
    stringstyle=\scriptsize\ttfamily\color[rgb]{0.627,0.126,0.941},
}

\makeatletter

\lstdefinelanguage{mlir}{
  morecomment = [l]{//},
  morestring=[b]", 
  sensitive = true,
  classoffset=0,
  classoffset=1, keywordstyle=\color{purple},
  morekeywords={
    soda.launch_func,
    soda.module,
    soda.return,
    func,
    define, declare, global, constant,
    internal, external, private,
    linkonce, linkonce_odr, weak, weak_odr, appending,
    common, extern_weak,
    thread_local, dllimport, dllexport,
    hidden, protected, default,
    except, deplibs,
    volatile, fastcc, coldcc, cc, ccc,
    x86_stdcallcc, x86_fastcallcc,
    ptx_kernel, ptx_device,
    signext, zeroext, inreg, sret, nounwind, noreturn,
    nocapture, byval, nest, readnone, readonly, noalias, uwtable,
    inlinehint, noinline, alwaysinline, optsize, ssp, sspreq,
    noredzone, noimplicitfloat, naked, alignstack,
    module, asm, align, tail, to,
    addrspace, section, alias, sideeffect, c, gc,
    target, datalayout, triple,
    blockaddress,
    return,
    step,
    func.func, scf.for, 
    linalg.matmul, linalg.generic,
    memref.load, memref.store, memref.subview,
    addi, muli, addf, mulf
  },
  classoffset=2, keywordstyle=\color{gray},
  morekeywords={
    memref, index, f32,
    !iarr_t, !varr_t,
    !AH_t, !AHW_t,
    !mr4x4_0, !mr4x4_1
  },
  alsoletter={\%, ., \!},
  keywordsprefix={\%},
}
\makeatother

%% file: IEEE/authors.tex
\author{\IEEEauthorblockN{Jude Haris\IEEEauthorrefmark{1},
Nicolas Bohm Agostini\IEEEauthorrefmark{2}\IEEEauthorrefmark{3}, \\
% Perry Gibson\IEEEauthorrefmark{3},
% Malith Jayaweera\IEEEauthorrefmark{2},
% Norm Rubin\IEEEauthorrefmark{2}, \\
Antonino Tumeo\IEEEauthorrefmark{2},
% José L. Abellán\IEEEauthorrefmark{4},
David Kaeli\IEEEauthorrefmark{3},
José Cano\IEEEauthorrefmark{1}
}
\IEEEauthorblockA{
\IEEEauthorrefmark{1}\textit{University of Glasgow},
Glasgow, Scotland, UK 
\IEEEauthorrefmark{2}\textit{Northeastern University},
Boston, MA, USA \\
% \IEEEauthorrefmark{4}\textit{University of Murcia},
% Murcia, Spain
\IEEEauthorrefmark{3}\textit{Pacific Northwest National Laboratory},
Richland, WA, USA
}}

%% file: lsts/baseline.tex
\begin{lstlisting}[style=customMLIR,language=mlir]
func.func @main (%input, %output) {
  // Allocation of %A and %B as outputs of other operation
  %A = memref.alloc () : () -> memref<60x80xfp32> 
  %B = memref.alloc () : () -> memref<80x72xfp32> 
  // ..Other operations happen...
  // %A, %B, and % are needed by the accelerator
  %C = memref.alloc () : () -> memref<60x72xfp32> 
  func.call @matmul_call(%A, %B, %C) // ...
}

func.func @matmul_call(...) {          
  // Declare constants (loop bounds and literals): %cX, ...
  accel.sendLiteral(%c0xFF) // reset opcode
  // Tiling by 4,4,4
  scf.for %m = %c0 to %c60 step %c4 { // first loop
    scf.for %k = %c0 to %c80 step %c4 { // second loop
      scf.for %n = %c0 to %c72 step %c4 { // innermost
        %sA = memref.subview %A[%m, %k] [4, 4] [1, 1]
        accel.sendLiteral(%c0x22)
        accel.send(%sA)
        %sB = memref.subview %B[%k, %n] [4, 4] [1, 1]
        %sC = memref.subview %C[%m, %n] [4, 4] [1, 1]
        accel.sendLiteral(%0x25)
        accel.send(%sB) 
        accel.recv {mode="accumulate"}(%sC)
  } } } return }
\end{lstlisting}

%% file: lsts/dma-allocation.tex
\begin{lstlisting}[style=customMLIR,language=mlir]
func.func @main (%input, %output) {
  // ...
  %A = memref.alloc () : () -> memref<60x80xfp32, #dma> 
  %B = memref.alloc () : () -> memref<80x72xfp32, #dma> 
  %C = memref.alloc () : () -> memref<60x72xfp32, #dma> 
  func.call @matmul_call(...)
  // ...
return } \end{lstlisting}

%% file: lsts/data-coalescing.tex
\begin{lstlisting}[style=customMLIR,language=mlir]
func.func @matmul_call(...) {          
  // Declare constants (loop bounds and literals): %cX, ...
  accel.sendLiteral(%c0xFF) : i32,i32->i32 // reset 
  scf.for %m = %c0 to %c60 step %c4 { // first loop
    scf.for %k = %c0 to %c80 step %c4 { // second loop
      scf.for %n = %c0 to %c72 step %c4 { // innermost
        %sA = memref.subview %A[%m, %k] [4, 4] [1, 1]
        %op0 = accel.load_opcode(%0x22)
        %sB = memref.subview %B[%k, %n] [4, 4] [1, 1]
        %sC = memref.subview %C[%m, %n] [4, 4] [1, 1]
        %op1 = accel.load_opcode(%0x25)
        accel.send([%op0,%sA,%sB,%op1])
        accel.recv {mode="accumulate"}(%sC)
  } } } return } \end{lstlisting}

%% file: lsts/pipeline.tex
\begin{lstlisting}[style=customMLIR,language=mlir]
func.func @matmul_call(...) {          
  // Declare constants (loop bounds and literals): %cX, ...
  accel.sendLiteral(%c0xFF) // reset opcode
  scf.for %m = %c0 to %c60 step %c4 { // first loop
    scf.for %k = %c0 to %c80 step %c4 { // second loop
      %sA = memref.subview %A[%m, %k] [4, 4] [1, 1]
      accel.sendLiteral(%c0x22)
      accel.send(%sA)
      %sB = memref.subview %B[%k, %c0] [4, 4] [1, 1]
      accel.sendLiteral(%0x25)
      accel.send(%sB) 
      scf.for %n = %c4 to %c68 step %c4 { // innermost
        %sA = memref.subview %A[%m, %k] [4, 4] [1, 1]
        accel.sendLiteral(%c0x22)
        accel.send(%sA)
        %n_last = arith.subi(%n, %c4)
        %sB = memref.subview %B[%k, %n] [4, 4] [1, 1]
        %sC = memref.subview %C[%m, %n_last] [4, 4] [1, 1]
        accel.sendLiteral(%0x25)
        accel.send(%sB) 
        accel.recv {mode="accumulate"}(%sC)
      }
      %sC = memref.subview %C[%m, %c68] [4, 4] [1, 1]
      accel.recv {mode="accumulate"}(%sC)
  } } return } \end{lstlisting}

%% file: main.bbl
\begin{thebibliography}{1}
\providecommand{\url}[1]{#1}
\csname url@samestyle\endcsname
\providecommand{\newblock}{\relax}
\providecommand{\bibinfo}[2]{#2}
\providecommand{\BIBentrySTDinterwordspacing}{\spaceskip=0pt\relax}
\providecommand{\BIBentryALTinterwordstretchfactor}{4}
\providecommand{\BIBentryALTinterwordspacing}{\spaceskip=\fontdimen2\font plus
\BIBentryALTinterwordstretchfactor\fontdimen3\font minus \fontdimen4\font\relax}
\providecommand{\BIBforeignlanguage}[2]{{%
\expandafter\ifx\csname l@#1\endcsname\relax
\typeout{** WARNING: IEEEtran.bst: No hyphenation pattern has been}%
\typeout{** loaded for the language `#1'. Using the pattern for}%
\typeout{** the default language instead.}%
\else
\language=\csname l@#1\endcsname
\fi
#2}}
\providecommand{\BIBdecl}{\relax}
\BIBdecl

\bibitem{agostini2023axi4mlir}
\BIBentryALTinterwordspacing
N.~B. Agostini, J.~Haris, P.~Gibson, M.~Jayaweera, N.~Rubin, A.~Tumeo, J.~L. Abellán, J.~Cano, and D.~Kaeli, ``{AXI4MLIR: User-Driven Automatic Host Code Generation for Custom AXI-Based Accelerators},'' 2023. [Online]. Available: \url{https://doi.org/10.48550/arXiv.2312.14821}
\BIBentrySTDinterwordspacing

\bibitem{Lattner2021mlir}
\BIBentryALTinterwordspacing
C.~{Lattner}, M.~{Amini}, U.~{Bondhugula}, A.~{Cohen}, A.~{Davis}, J.~{Pienaar}, R.~{Riddle}, T.~{Shpeisman}, N.~{Vasilache}, and O.~{Zinenko}, ``{MLIR: Scaling Compiler Infrastructure for Domain Specific Computation},'' in \emph{IEEE/ACM International Symposium on Code Generation and Optimization}, ser. CGO'21.\hskip 1em plus 0.5em minus 0.4em\relax Seoul, Korea (South): IEEE, 2021, pp. 2--14. [Online]. Available: \url{https://doi.org/10.1109/CGO51591.2021.9370308}
\BIBentrySTDinterwordspacing

\bibitem{shabani2023hpca}
\BIBentryALTinterwordspacing
H.~Shabani, A.~Singh, B.~Youhana, and X.~Guo, ``Hirac: A hierarchical accelerator with sorting-based packing for spgemms in dnn applications,'' in \emph{IEEE International Symposium on High-Performance Computer Architecture}, ser. HPCA'23.\hskip 1em plus 0.5em minus 0.4em\relax Montreal, QC, Canada: IEEE, 2023, pp. 247--258. [Online]. Available: \url{https://doi.org/10.1109/HPCA56546.2023.10070977}
\BIBentrySTDinterwordspacing

\bibitem{kim2023hpca}
\BIBentryALTinterwordspacing
B.~Kim, S.~Li, and H.~Li, ``Inca: Input-stationary dataflow at outside-the-box thinking about deep learning accelerators,'' in \emph{IEEE International Symposium on High-Performance Computer Architecture}, ser. HPCA'23.\hskip 1em plus 0.5em minus 0.4em\relax Montreal, QC, Canada: IEEE, 2023, pp. 29--41. [Online]. Available: \url{https://doi.org/10.1109/HPCA56546.2023.10070992}
\BIBentrySTDinterwordspacing

\bibitem{hsia2023asplos}
\BIBentryALTinterwordspacing
S.~Hsia, U.~Gupta, B.~Acun, N.~Ardalani, P.~Zhong, G.-Y. Wei, D.~Brooks, and C.-J. Wu, ``Mp-rec: Hardware-software co-design to enable multi-path recommendation,'' in \emph{Proceedings of the 28th ACM International Conference on Architectural Support for Programming Languages and Operating Systems, Volume 3}, ser. ASPLOS 2023.\hskip 1em plus 0.5em minus 0.4em\relax New York, NY, USA: Association for Computing Machinery, 2023, p. 449–465. [Online]. Available: \url{https://doi.org/10.1145/3582016.3582068}
\BIBentrySTDinterwordspacing

\bibitem{munoz2023asplos}
\BIBentryALTinterwordspacing
F.~Mu\~{n}oz Mart\'{\i}nez, R.~Garg, M.~Pellauer, J.~L. Abell\'{a}n, M.~E. Acacio, and T.~Krishna, ``Flexagon: A multi-dataflow sparse-sparse matrix multiplication accelerator for efficient dnn processing,'' in \emph{Proceedings of the 28th ACM International Conference on Architectural Support for Programming Languages and Operating Systems}, ser. ASPLOS'23.\hskip 1em plus 0.5em minus 0.4em\relax New York, NY, USA: Association for Computing Machinery, 2023, p. 252–265. [Online]. Available: \url{https://doi.org/10.1145/3582016.3582069}
\BIBentrySTDinterwordspacing

\bibitem{Rao2018sdc}
\BIBentryALTinterwordspacing
Q.~Rao and J.~Frtunikj, ``Deep learning for self-driving cars: Chances and challenges,'' in \emph{IEEE/ACM 1st International Workshop on Software Engineering for AI in Autonomous Systems}, ser. SEFAIS'18.\hskip 1em plus 0.5em minus 0.4em\relax New York, NY, USA: Association for Computing Machinery, 2018, p. 35–38. [Online]. Available: \url{https://doi.org/10.1145/3194085.3194087}
\BIBentrySTDinterwordspacing

\bibitem{mlir2020linalg}
\BIBentryALTinterwordspacing
M.~Developers, ``{'linalg' Dialect},'' 2020, online accessed on 11-04-2023. [Online]. Available: \url{https://mlir.llvm.org/docs/Dialects/Linalg/}
\BIBentrySTDinterwordspacing

\end{thebibliography}
